\documentstyle[emulateapj]{article}

\lefthead{Armitage \& Natarajan}
\righthead{Lense-Thirring precession of accretion disks}

\begin{document}

\submitted{ApJ, in press}

\title{Lense-Thirring precession of accretion disks around compact objects}

\author{Philip J. Armitage and Priyamvada Natarajan}
\affil{Canadian Institute for Theoretical Astrophysics, McLennan Labs,
	60 St George St, Toronto, M5S 3H8, Canada\\ email: armitage@cita.utoronto.ca, 
	priya@cita.utoronto.ca}

\begin{abstract}
Misaligned accretion disks surrounding rotating compact objects experience a torque 
due to the Lense-Thirring effect, which leads to precession of the inner 
disk. It has been suggested that this effect could be responsible for 
some low frequency Quasi-Periodic Oscillations observed in the X-ray 
lightcurves of neutron star and galactic black hole systems. We investigate 
this possibility via time-dependent calculations of the response of the inner 
disk to impulsive perturbations 
for both Newtonian point mass and Paczynski-Wiita potentials, and compare the results 
to the predictions of the linearized twisted accretion disk equations. For 
most of a wide range of disk models that we have considered,
the combination of differential 
precession and viscosity causes the warps to decay extremely rapidly. Moreover, 
at least for relatively slowly rotating objects, linear calculations in a Newtonian 
point mass potential provide a good measure of the damping rate, provided only
that the timescale for precession is much shorter than the viscous 
time in the inner disk. The typically rapid decay rates suggest that coherent 
precession of a fluid disk would not be observable, though it remains 
possible that the damping rate of warp in the disk could be low enough to 
permit weakly coherent signals from Lense-Thirring precession.
\end{abstract}	


\noindent{\em Animation of warp evolution is available at:
{\tt http://www.cita.utoronto.ca/$^\sim$armitage/lense{\underline{ }}thirring.html}}          

\section{INTRODUCTION}

Tilted orbits about a rotating object experience a torque due to 
the general relativistic Lense-Thirring effect (Lense \& Thirring 1918), 
which causes the plane of the orbit to precess. For a fluid accretion 
disk, the differential precession with radius causes stresses and 
dissipation. If the torque is strong enough compared to the internal 
viscous forces, the result is that the inner regions of the disk 
are forced to align with the spin of the central neutron star or
black hole (Bardeen \& Petterson 1975). 

This process has several important effects. If the angular momentum 
of disk material at large radius is misaligned with respect to the 
rotation axis of the central object, then alignment of the inner regions 
by the Bardeen-Petterson effect implies that the disk must possess 
a large-scale warped or twisted shape. It is then obvious that this 
will both modify the emergent spectrum, and determine the direction of 
jets accelerated from the inner disk region. The consequences of this
for disks and jets in Active Galactic Nuclei are well known, and have
been discussed for some time (Bardeen \& Petterson 1975; Rees 1978; 
Scheuer \& Feiler 1996; Natarajan \& Pringle 1998; Natarajan \& Armitage 1999).

More recently, it has been suggested that Lense-Thirring precession may be 
{\em directly} observable in the form of Quasi-Periodic Oscillations (QPOs) 
in the X-ray lightcurves of low-mass X-ray binaries (Stella \& Vietri 1998), 
including galactic black hole candidates (Cui, Zhang \& Chen 1998). If confirmed, 
this identification promises to provide constraints on the equation of 
state of material at nuclear densities, and on the spin parameter of 
stellar mass black holes. It would also constitute one of the few 
arenas where the Lense-Thirring effect might be detectable -- the 
measurement of this precession for orbits around the Earth 
(Ciufolini et al. 1998) is, itself, controversial. However, 
other explanations for the observed QPOs remain viable.
For neutron stars especially, the possibilities for interaction 
between the disk and a magnetosphere are legion, and in general 
remain poorly understood (see e.g. Miller \& Stone 1997; Spruit \& 
Taam 1993; Ghosh \& Lamb 1979). There are also other relativistic effects possible 
for purely flat, fluid disks around black holes, which can produce 
oscillations with frequencies in the right range (e.g. Nowak et al. 1997).

Estimates of the frequency of Lense-Thirring precession for parameters
appropriate to neutron stars are, at least to an order of magnitude, 
in agreement with the observed frequencies of some QPOs (Stella \& Vietri 1998). 
The main theoretical uncertainty in applying the model is then the damping rate 
of perturbations excited in the inner disk. This was computed by 
Markovic \& Lamb (1998), who considered the global modes of the disk 
described by the linearized version of the twisted disk evolution 
equations. They found that the combination of differential precession and 
disk viscosity was generally highly deleterious to the survival of warps 
in the disk. The damping rates they obtained were sufficiently rapid 
as to cast serious doubt on the Lense-Thirring interpretation of QPOs, 
although some relatively weakly damped modes 
were found when the torques from radiation were also included. In response, 
Vietri \& Stella (1998) suggested that damping of vertical motions
might be much weaker than that assumed by Markovic \& Lamb (1998), 
due either to a radically modified disk structure, or a supposed 
more general suppression of damping in the inner, aligned portions
of the disk. 

In this paper, we compute numerically the response of a viscous disk 
to finite amplitude perturbations in the innermost regions where the 
Lense-Thirring torque is strong. Following Markovic \& Lamb (1998), 
we employ the equations derived by Papaloizou \& Pringle (1983) that 
describe the hydrodynamic response of a thin, viscous disk that evolves 
essentially diffusively. Our approach is to commence with a steady, 
flat disk, perturb the inner disk impulsively to generate a warp, 
and then evolve the full twisted disk equations to follow the warp 
as it precesses and decays. We compare our results both with 
the linear evolution investigated previously, and with the evolution 
in a Paczynski-Wiita (1980) potential that models some of the non-Newtonian 
effects expected close to neutron stars and black holes.

The outline of the paper is as follows. We set out the equations solved 
in $\S$2, and estimate the parameters appropriate for accretion disks 
in X-ray binaries. In $\S3$ we present results from a grid of models with varying 
viscosity laws, potentials, and approximations to the twist equation. 
$\S$4 summarizes our conclusions.

\section{EVOLUTION EQUATIONS FOR A TWISTED DISK}

The evolution of a geometrically thin, warped accretion disk in the diffusive 
regime can be described using the equations developed by Papaloizou \& 
Pringle (1983).
In terms of the angular momentum density of the disk ${\bf L}(R,t)$, the 
governing equation is (Pringle 1992),
\begin{eqnarray}
 { {\partial {\bf L}} \over \partial t } & = & 
 {1 \over R } {\partial \over {\partial R}}
 \left[ {{ (\partial / \partial R) [ \nu_1 \Sigma R^3 (-\Omega^{'}) ] } \over 
 { \Sigma (\partial / \partial R) (R^2 \Omega) }} {\bf L} \right] \nonumber \\
 & + & {1 \over R } {\partial \over {\partial R}}
 \left[ {1 \over 2} \nu_2 R \vert {\bf L} \vert { {\partial {\bf l}} \over 
 {\partial R} } \right] \nonumber \\
 & + & {1 \over R } {\partial \over {\partial R}} 
 \left\{ \left[ { { {1 \over 2} \nu_2 R^3 \Omega \vert \partial {\bf l} / 
 \partial R \vert^2 } \over { (\partial / \partial R) (R^2 \Omega)} } + 
 \nu_1 \left( { {R \Omega^{'}} \over \Omega } \right) \right] {\bf L} \right\} \nonumber \\
 & + & {\bf \Omega_p} \times {\bf L}.
\label{eq1}
\end{eqnarray} 
Here ${\bf L} = (GM)^{1/2} \Sigma R^{1/2} {\it \hat l}$, $\Sigma$ is the
surface density, and ${\vec l}$ is a unit vector normal to the disk 
surface. The angular velocity in the disk at radius $R$ is $\Omega(R)$, 
and $\Omega^{'} \equiv {\rm d} \Omega / {\rm d} R$. We consider Keplerian 
orbits in two possible forms for the potential, a Newtonian point mass 
potential, 
\begin{equation} 
 \psi = { {-GM} \over R} 
\label{eq1b}
\end{equation}
and a pseudo-Newtonian potential (Paczynski \& Wiita 1980),
\begin{equation}
 \psi = { {-GM} \over {R-R_g} }
\label{eq1c}
\end{equation}
where $R_g = 2 GM / c^2$. The latter potential reproduces correctly some 
features of orbits around compact objects -- for example the existence and 
location of an innermost stable orbit -- though we include it here mainly 
to gauge the sensitivity of equation (\ref{eq1}) to changes in the angular 
velocity profile. $\nu_1$ and $\nu_2$ are ``viscosities''
corresponding to the azimuthal and vertical shear respectively, and the 
last term on the right hand side of the equation is the Lense-Thirring torque.
For a black hole with angular momentum ${\bf J}$, for example, 
${\bf \Omega_p} = {\bf \omega_p} / R^3 = 2 {\bf J} G / c^2 R^3$.  
We note that this fundamentally Newtonian approach is appropriate only 
for black holes with modest values of the spin parameter. In particular, 
the behavior of disks around rapidly rotating, Kerr black holes, is formally 
not addressed by these calculations. Although there are no direct 
measurements of black hole spins, observations are suggestive 
of a broad range of spin parameters for black holes in X-ray 
binaries (Zhang, Cui \& Chen 1997). 

Equation (\ref{eq1}) represents a vast simplification to the general 
equations for the time dependent evolution of a warped accretion 
disk, which in general require a full hydrodynamic treatment 
(see e.g. the simulations of Larwood et al. 1996). The conditions 
under which equation (\ref{eq1}) is valid have been investigated 
in detail (Papaloizou \& Pringle 1983; Papaloizou \& Lin 1994; 
Ogilvie 1999), with the result that diffusive evolution is likely 
to apply to thin disks in Active Galactic Nuclei and, probably,  
X-ray binaries (Pringle 1999), which are the systems of interest here. 
Conversely wave-like evolution is likely to dominate for disks around young 
stars. 

Equation (\ref{eq1}) can be derived either from consideration of mass and
angular momentum conservation of two neighboring annuli in the disk sharing their
angular momentum via viscosity (Papaloizou \& Pringle 1983; Pringle 1992), 
or via a hydrodynamic treatment of the warped disk valid in the linear 
regime (Papaloizou \& Pringle 1983). More recent analysis (Ogilvie 1999) has 
shown that an equation of this type is valid even for strongly warped disks, 
though, in general, {\em neither} $\nu_1$ nor $\nu_2$ so defined are equal 
to the usual kinematic viscosity $\nu$ of the Navier-Stokes equation and 
planar disk theory. In fact, the relation between these quantities for an 
arbitrary warp is extremely complex (Ogilvie 1999). For this paper, we 
will take the ratio of $\nu_1$ to $\nu_2$ to be a free parameter, and 
examine the influence of this choice on the decay of warps in the inner 
disk. Clearly, models with low $\nu_2$ viscosity are the most favorable 
for sustaining observable precession of the inner disk regions. 

There are two relevant timescales in this problem, the precession 
time and the viscous timescale corresponding to the vertical shear, 
which we define as,
\begin{equation} 
 t_{\rm prec} \equiv { {2 \pi} \over {\vert {\bf \Omega_p} (R) \vert}} \,\,\,\,\,\,\,\,\
 t_{\nu_2} \equiv { R^2 \over {\nu_2 (R)} }.
\label{eq3}
\end{equation} 
The balance between these timescales determines whether the 
Bardeen-Petterson effect is able to align the inner disk with the 
spin of the central object. Roughly speaking, this will occur 
if $t_{\rm prec} \lesssim t_{\rm \nu_2}$ at $R=R_{\rm in}$ 
(Kumar \& Pringle 1985; Scheuer \& Feiler 1996; Natarajan \& Armitage 1999). 

This ratio of timescales, $t_{\rm prec} / t_{\nu_2}$, can be 
estimated as follows. For definiteness, we assume that the disk 
surrounds a black hole with spin parameter $a$, for which 
\begin{equation} 
 \omega_p = 2 a c \left( {GM} \over c^2 \right)^2,
\label{eq3b}
\end{equation}
and the corresponding precession timescale is,
\begin{equation} 
 t_{\rm prec} = { {\pi R^3} \over {2 a c} } \left( {c^2 \over {GM}} \right)^2.
\label{eq3c}
\end{equation}
If the $\nu_1$ disk viscosity is parameterized via the Shakura-Sunyaev (1973) 
prescription, $\nu_1 = \alpha c_s^2 / \Omega$, where $\alpha$ is a dimensionless 
parameter and $c_s$ is the local sound speed, then $t_{\nu_1} = R^2 / \nu_1$ is 
given by,
\begin{equation} 
 t_{\nu_1} = {1 \over {\alpha \Omega} } \left( R \over H \right)^2,
\label{eq3d}
\end{equation} 
where $H$ is the disk scale height. We then have,
\begin{equation}
  { t_{\rm prec} \over t_{\nu_1} } \simeq 10^{-2} 
  \left( \alpha \over 0.01 \right) 
  \left( a \over 0.5 \right)^{-1}   
  \left( R \over {3 R_g} \right)^{3/2} 
  \left( { H / R } \over 0.1 \right)^2
\label{eq3e}
\end{equation} 
where we have adopted rough estimates for $\alpha$ and $H/R$. Obviously, there are 
large variations possible in all of these parameters, for example reducing 
$H/R$ to $10^{-2}$  
would decrease $t_{\rm prec} / t_{\nu_1}$ by two orders of magnitude. However, 
if $\nu_2 \sim \nu_1$, then this simple estimate suggests that 
the precessional timescale in the inner disk is likely to be perhaps two or three 
orders of magnitude shorter than the local viscous timescale. Accordingly, 
we choose parameters for our calculations that match this regime, and the 
a priori more favorable one (for observing disk precession) where the 
viscosity, and hence the damping of warps, is weaker.

We solve equation (\ref{eq1}) using the explicit finite difference
method described by Pringle (1992), with zero torque boundary 
conditions at $R_{\rm in}$ and $R_{\rm out}$. The outer boundary 
condition is applied at a radius large enough as to not affect the 
inner disk evolution over the time period of interest. Typically, 
$R_{\rm out} / R_{\rm in} = 16$ suffices. Moderately high numerical 
resolution is needed to resolve the strongly twisted disk configurations
that develop, we use 400 to 800 grid points logarithmically spaced between 
$R_{\rm in}$ and $R_{\rm out}$. Tests show that this resolution is
more than adequate.

For our calculations we adopt units in which $R_{\rm in} = 1$, and 
$\vert \Omega_p (R_{\rm in}) \vert = 8$. We take the viscosities to be 
power laws in radius, $\nu_1 = \nu_{10} R^\delta$, $\nu_2 = \nu_{20} R^\delta$, 
and consider cases where $\nu_{20} / \nu_{10} = 5, 1, 0.2$. We consider an 
initially aligned, planar disk, apply a simple warp perturbation to the 
inclination,
\begin{equation} 
  \Delta i = \Delta i_0 \exp^{-(R-R_{\rm p})^2 / \Delta R_{\rm p}^2},
\label{eq4}   
\end{equation}  
and follow the perturbed disk as the warp decays and precesses. We 
note that this form of perturbation, in which there is initially no twist in 
the disk warp, generally decays {\em slower} than a twisted 
configuration for which the radial scale on which components of 
$\vec{l}$ change is smaller.

These initial conditions are appropriate if the outer accretion disk is 
aligned with the equatorial plane of the spinning black hole. Alternatively, 
the outer disk could be misaligned, either as a consequence of warping 
instabilities (Pringle 1996; Maloney \& Begelman 1997; Schandl \& Meyer 1994), 
or because the black hole retains an initially misaligned angular 
momentum vector (King \& Kolb 1999). However, even in these cases, we 
expect that the inner disk will be aligned with the spin of the hole 
by the Bardeen-Petterson (1975) effect. For standard models 
of the disk viscosity, this alignment radius is large compared to the 
region of the disk from which QPOs are expected to originate. This 
conclusion is reasonably robust, at least for steady disks, 
since it follows simply from the relative timescale for precession 
versus viscous evolution. Furthermore, we show later that models 
in which alignment would {\em not} occur (those with relatively large 
values of the disk viscosity), are those in which perturbations damp 
most rapidly.  

\section{RESULTS} 

\subsection{Newtonian point mass potential}

Fig.~1 shows the decay of a warp in the inner regions of the 
disk. The parameters of the initial perturbation were: 
$R_{\rm p} = 2$, $\Delta R_{\rm p} = 0.2$, and $\Delta i_0 = 0.1$. 
The azimuthal angle of the ascending node, $\gamma$, was initially constant 
with radius, i.e. the initial state had no twist. The time slices
are plotted at intervals of $\Delta t = 5$, which is a little
less than the precession time, $t_{\rm prec} = 2 \pi$, at $R_{\rm p}$.
The disk viscosity was taken to be $\nu_1 = \nu_2 = 10^{-2} R^{3/2}$, 
which gives $t_{\rm prec} / t_{\nu_2} = 4 \times 10^{-2}$. 

From the Figure, it can be seen that the warp decays rapidly, with 
the peak inclination decaying by roughly an order of magnitude per
precession time. As it does so, the warp diffuses radially, and the 
differential precession with radius induces a strong twist in the 
disk shape. This is shown in Fig.~2, which plots the shape of the 
disk at the instant when the perturbation is applied and after 
a time $\Delta t = 15$, around two precession timescales later. The 
warp becomes increasingly twisted with time, which reduces the 
characteristic radial lengthscales for variations in ${\vec l}$, and 
thereby shortens the timescale ($\sim \Delta R^2 / \nu_2$) 
required for diffusion to flatten the disk.

Fig.~3 shows the temporal behavior of the warp for a grid of 
disk models with varying assumed forms for the viscosity.
We consider viscosity laws with $\delta = 3/2$ and $\delta = 0$, 
for a range of $\nu_{10}$ and $\nu_{20}$. 

For the cases where the viscosity is weak (i.e. low $\nu_{20}$), 
the decay of the warp depends almost exclusively on the value 
of $\nu_{2}$ for the range of $\nu_{1}$ considered here. The initial 
rate of decay is found to be proportional to $\nu_2^{-1}$, but 
this phase lasts for at most a few precession times. Subsequently, 
the disk inclination is found to decline more rapidly with time, 
so that at the end of the runs the rate of decay ${\rm d} \log i / {\rm d}t$ 
is similar for the various choices of $\nu_{2}$. The runs with 
differing $\delta$ are very similar, the main distinction being 
the trivial one that the $\delta = 0$ calculations have lower 
viscosity at $R=R_{\rm p}$ due to our choice of normalization. 
Thus, the decay rates are correspondingly lower.

Our strongest viscosity runs show some dependence of the damping 
rate on $\nu_{1}$ as well as $\nu_{2}$. The model with large 
$\nu_{1} / \nu_2$ exhibits substantially faster decay. This is 
due to the larger radial velocity advecting the perturbation 
inwards, which causes a more rapid decline in the inclination at
the fixed radius $R_{\rm p}$. 

\subsection{Precession rate}

Fig.~4 shows the precession rate, ${\rm d}\gamma / {\rm d}t$, for the 
annulus at $R=2$ in the $\delta = 3/2$ models. The Lense-Thirring 
precession rate at this radius for our choice of units corresponds
to $({\rm d}\gamma / {\rm d}t) = 1$, and this is the initial rate 
of precession for the warped disk in all the models. Thereafter, 
the precession rate declines as the increasingly twisted disk 
generates internal viscous torques that oppose further distortion 
of the disk shape, with the decay rate of the precession varying with $\nu_2$ in the 
same manner as the disk inclination. Comparing Figs.~3 and 4, we
find that the precession of the disk annulus remains within 
10\% of its initial value up to the epoch when the disk inclination 
has decayed by roughly an order of magnitude. For all the models 
considered here this is a short timespan, of the order of a few
precession periods.

Converting these results to a prediction of the coherence (or lack 
thereof) of a Lense-Thirring feature in the power spectrum of X-ray 
binaries would require detailed knowledge of the emission mechanisms
producing the X-ray flux, which is generally lacking. However, it is 
clear that for the models considered here, the perturbed disk is 
able to maintain a coherent precessional motion, at a fixed radius, 
for at most a few precession periods. This might be sufficient to 
generate weakly coherent oscillations in the disk emissivity (and 
some of the observed QPOs are indeed of this type), but 
would be unable to lead to oscillations with a high quality factor.
Moreover, this estimate 
most probably overstates the coherence of the X-ray signal, since 
any plausible emission mechanism will sample a range of radii
which will have different precession frequencies. 

\subsection{Comparison with linearized equations}

It is of interest to compare our results with the comprehensive analysis 
of the linearized twist equations, including Lense-Thirring precession, presented 
by Markovic \& Lamb (1998). For small warp amplitudes, where $\beta \ll 1$ and 
$R \beta^{'} \ll 1$ (here $\beta$ is the local angle of tilt), 
the surface density of the disk remains a fixed 
function of radius, and the warp can be completely described by two components 
of the local tilt vector, $l_x$ and $l_y$. This simplifies equation (\ref{eq1}) 
significantly, and the resulting equation can be expressed in terms of 
a single complex quantity $W = \beta e^{i \gamma}$. For $\nu_2 \propto R^{3/2}$ 
the evolution of $W$ is described by,
\begin{equation}
  { {\partial W} \over {\partial t} } = 
  { \omega_p \over R^3 } i W + {\nu_{20} \over 2} 
  { \partial \over {\partial R} } \left( 
  R^{3/2} { {\partial W} \over {\partial R} } \right).
\label{eq10}
\end{equation}  
Markovic \& Lamb (1998) proceeded to solve for the modes of this equation, and 
the corresponding equation including terms representing radiative torques. For our 
purposes of comparing results with 
the previous calculations, we instead solve equation (\ref{eq10}) as an 
initial value problem with the initial conditions defined by equation 
(\ref{eq4}). This can straightforwardly be done using identical methods 
to those employed for the full evolution equation.

Figure 5 shows the comparison between the disk evolution computed with equations 
(\ref{eq1}) and (\ref{eq10}). Two choices of viscosity are used, the $\delta=3/2$, 
$\nu_{10}=\nu_{20}=10^{-2}$ model discussed already, and a model where the 
viscous time is an order of magnitude longer. Considering this second case 
first, for these parameters, where $t_{\rm prec} \ll t_{\nu}$, the decay of 
the warp is essentially identical between the two 
calculations, both in the rate and the detailed warp shape. 
This regime is the only one where there is some possibility of precession 
surviving for long enough to be potentially observable, and for warps of 
this (relatively modest) amplitude we therefore conclude that the linear approach 
provides an excellent approximation to the evolution described by 
equation (\ref{eq1}). For the model where $t_{\rm prec}$ and $t_{\nu}$ are 
more comparable (the upper panel in Fig.~5), there are some differences 
between the detailed shape of the warp in the two calculations. These 
are due to changes in the surface density of the inner disk in response 
to the warp, which are ignored in the linear treatment. Even in 
this case though the {\em decay rates} are found to be very similar, 
and there is certainly no evidence that including the extra terms 
in equation (\ref{eq1}) can cause significantly slower decay than 
is obtained in the linear regime.

\subsection{Paczynski-Wiita potential}

A rather analogous situation to the preceeding Section is found for 
the comparison between the Newtonian point mass and 
Paczynski-Wiita potentials. Figure 6 shows the evolution of disk 
inclination as a function of radius for the two potentials, using the 
same parameters as for Fig.~1. There are significant differences 
between the shape of the warp in the two calculations. At late 
times the warp in a Paczynski-Wiita potential sustains a higher 
amplitude at the inner disk edge than in the Newtonian case, 
and in principle this would be more favorable for producing an 
observable signal of disk precession. This is attributable 
to the lower surface density (and correspondingly more rapid radial 
inflow for a given disk viscosity) at small radii in the non-Newtonian 
potential. However, the damping rates for the two calculations 
differ only to a qualitatively insignificant degree -- for this 
choice of parameters both warps decay extremely rapidly. Moreover, 
for less viscous disks, where the warps decay more slowly, the 
differences between the evolution in the two potentials quickly 
become much smaller. 

\subsection{Radiation induced warping}

In the preceeding discussion, we have investigated the {\em damping} 
rates of warps in disks whose evolution is driven solely 
by induced precession and internal disk viscosity.
The influence of radiation forces caused by re-emission of 
radiation from a central source of luminosity is known 
to lead to {\em growing} warping modes of the disk (Pringle 1996) in 
some circumstances, so it is of interest to consider how 
they may alter the evolution of the inner disk.
The first point to note is that the large scale warping modes 
discussed by Maloney, Begelman \& Pringle (1996; see also 
Maloney \& Begelman 1997; Maloney, Begelman \& Nowak 1998; 
Wijers \& Pringle 1999) are probably irrelevant to the 
discussion. These modes grow and precess on timescales 
that are of the order of the viscous timescale of the outer 
edge of the disk, typically tens to hundreds of days for 
X-ray binaries. The most promising circumstances for 
observing Lense-Thirring precession require $t_{\rm prec} \ll t_{\nu_2}$ 
in the inner disk, in which case any outer warp will be flattened 
into the equatorial plane well outside the innermost regions.
However, the low surface density of the inner regions, if they 
remain optically thick, allow for the possibility that radiation 
reaction forces could play an additional significant role there. 
Indeed, consideration of the linearized twist equations, including 
the radiative torque (Markovic \& Lamb 1998), suggests that the decay 
rate of some twisted
disk modes could be substantially reduced over the case where radiation 
is neglected.

Although highly suggestive, we feel that the influence of radiation 
reaction forces on the survival of inner disk precession is likely 
to be considerably more subtle, owing primarily to the importance of shadowing.
Shadowing of parts of the outer disk by twists at smaller radii is 
an inherently non-linear effect that provides additional coupling
between different radii in the disk. For a disk illuminated by a
point source at the origin, it modifies the growth rate even for 
negligible disk inclinations, where the linear approximation would 
otherwise be extremely accurate. We have found that the sense of 
the modification is typically to reduce the growth rate for
disks that are subject to radiation warping at large radius, and 
we anticipate that there would be equally significant effects for 
the much more tightly wound modes present at small radii when the 
effects of Lense-Thirring precession are included.
Additionally, at radii of just a few $R_g$, the assumption of a central 
illuminating point source (which is an excellent approximation for 
the usual radiation driven modes at very large disk radii) breaks 
down, and the real radiation field is likely to be much more complex.
It is unclear to what degree these effects alter the damping rates 
computed by Markovic \& Lamb (1998), but clearly further investigation 
in this area is merited.

\section{DISCUSSION}

In this paper, we have presented results from a time-dependent 
treatment of the evolution of perturbations in the inner regions 
of accretion disks which are aligned with the spin axis of the 
accretor by the Bardeen-Petterson effect. We find that such 
perturbations decay rapidly. For a weak viscosity 
(one for which the viscous timescale corresponding to $\nu_2$ 
is much greater than the local precession timescale), 
the inner parts of the disk would be expected to be accurately aligned 
with the rotation axis. An impulsively generated warp then decays
at a rate that depends solely on the viscosity acting on the $(R,z)$ stress, 
conventionally parameterized as $\nu_2$ (Papaloizou \& Pringle 1983). The 
initial damping rate is proportional to $\nu_2$, but 
after a few precession periods the damping is greatly 
enhanced as a consequence of the strong differential 
precession and twisting of the disk shape. This twisting 
rapidly reduces the radial lengthscale over which disk 
viscosity must act in order to flatten the disk into the 
equatorial plane, and leads to rapid damping.  

Our calculations are based on a numerical treatment of the 
full twisted disk equations derived by Papaloizou \& Pringle (1983), 
which can readily be extended to arbitrary rotation laws (Pringle 1992). We 
have compared the evolution of warps in potentials arising from 
a Newtonian point mass, and a Paczynski-Wiita potential that mimics 
some features of hydrodynamics in a Schwarzschild metric. For 
disk parameters chosen such that warps are at least relatively 
long lived, we find that the decay rate of warps in these two 
potentials are very similar, and accurately described by a linearized 
treatment of the twisted disk equations. Disks where the inflow 
velocity is larger exhibit some qualitative differences, both 
between the two potentials and when compared to the linear results, 
but these disks show even faster damping than expected on the basis 
of the value of $\nu_2$ alone. They are not favorable cases for 
producing long lived, observable, precession.
Advection dominated flows (e.g. Narayan \& Yi 1994), where the 
radial inflow velocity is typically a significant fraction 
of the orbital velocity, would be extreme examples of this 
effect. Such disks seem even less likely to be able to support 
coherent Lense-Thirring precession. 

These calculations, and those of Markovic \& Lamb (1998), assume that 
the accreting material in the vicinity of the black hole can be treated as 
a fluid accretion disk. This would be wrong if the inner disk was prone 
to breaking into blobs, for example via the development of the Lightman-Eardley (1974) 
or other instabilities (e.g. Ghosh 1998). Several authors have suggested 
that the inner disk in X-ray binaries is, indeed, susceptible to such 
instabilities (e.g. Taam \& Lin 1984; Lasota \& Pelat 1991; Cannizzo 1996). 
The behavior of accreting gas in such a model could be very different from 
that considered here, and is beyond the scope of this paper. Further study 
of such instabilities, in the context of currently favored magnetohydrodynamic 
models for the origin of the disk viscosity, would be valuable.  

The strength of the viscosity in planar accretion disks is 
constrained, albeit poorly, by a variety of observational 
methods (see e.g. Cannizzo, Chen \& Livio 1995). By contrast, there
are essentially no constraints whatsoever on how efficiently a
disk can damp warp, especially at the small radii that are of 
relevance to Lense-Thirring precession (at large radii weak 
constraints are possible from observing which systems may be 
prone to radiation induced warping). This introduces a major 
uncertainty in any attempt to apply the results to interpreting 
the possible origins of QPOs in X-ray binaries. However, if 
$\nu_1 \sim \nu_2$, it seems clear that warps in the inner disk 
are damped on a timescale of the order of a single precession 
period, or less. This is the same result as was found by 
Markovic \& Lamb (1998) from a linear analysis, and it suggests 
that precession is unlikely to be the origin of low-frequency 
QPOs in disks around neutron stars and galactic black hole candidate 
sources. The important caveat is that we know of no definite reason, either 
from observational constraints or from theoretical arguments, 
why $\nu_2$ might not be very much smaller that $\nu_1$ in 
the inner disk. In this regime, differential precession still 
leads to strong twisting and eventual rapid decay of the warp. 
However, our lowest $\nu_2$ runs maintain a reasonable warp amplitude 
for $\sim 10$ precession periods, along with a roughly constant 
precession rate. This could lead to at least weakly coherent 
signals that have as their origin Lense-Thirring precession. 
Better theoretical understanding of the hydrodynamics of disk 
warps will be needed to investigate this possibility. Recent work 
(Ogilvie 1999) has made progress towards this goal.

\acknowledgements

We thank Jim Pringle for making available his numerical disk 
evolution code, Gordon Ogilvie for sharing results in advance
of publication, and the referee Wei Cui for a prompt and very 
helpful report.

\newpage

\begin{figure}
\plotone{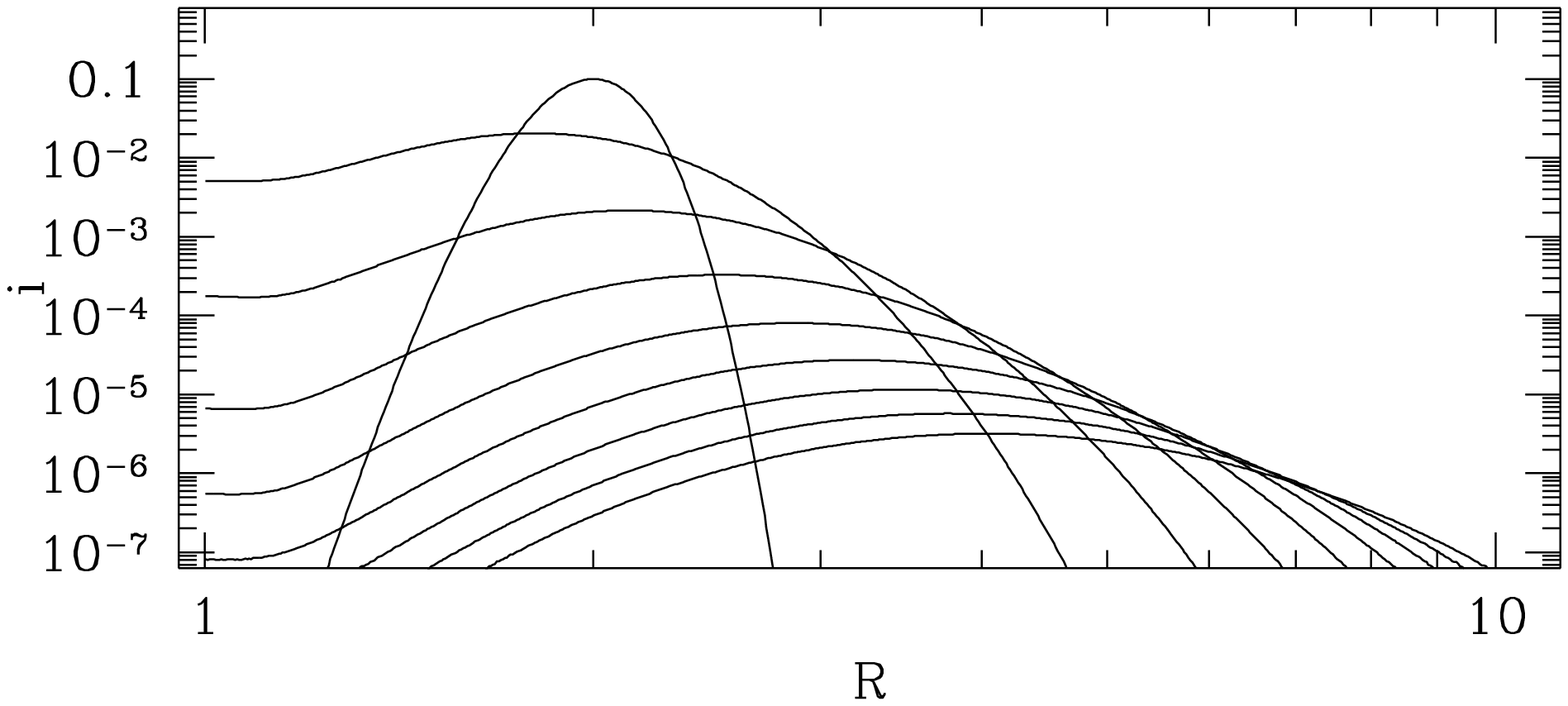}
\caption{Decay of the warp for the run with $\nu_1 = \nu_2 = 10^{-2} R^{3/2}$. 
         From top downwards, time slices plot the inclination angle 
         in radians at intervals of 
         $\Delta t = 5$. The time units are such that 
         $\vert {\bf \Omega_p} \vert^{-1} = 1$ at $R=2$.}
\end{figure}       
            
\begin{figure}
\plotone{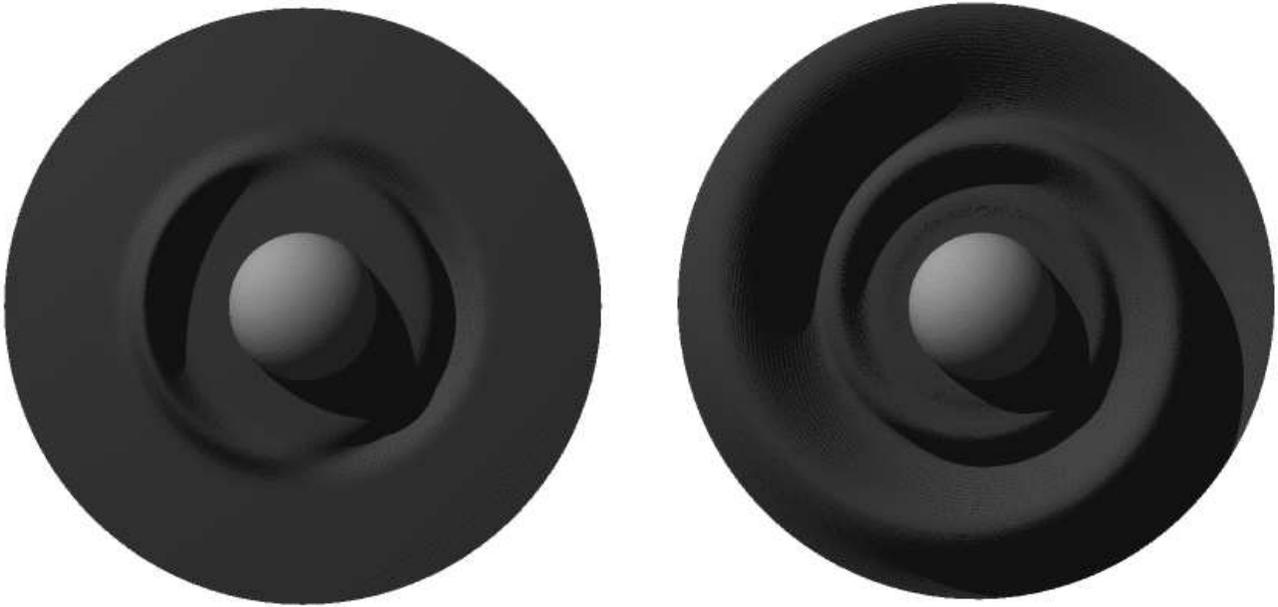}
\caption{Disk shape for the run with $\nu_1 = \nu_2 = 10^{-2} R^{3/2}$. The 
	 left image shows the initial warp perturbation at $t=0$, the right 
	 image the warp at $t=15$. Note that the {\it amplitude} of the 
	 warp has been vastly exaggerated in the second image to show 
	 the shape of the rapidly decaying perturbation.}
\end{figure}  

\begin{figure}
\plotone{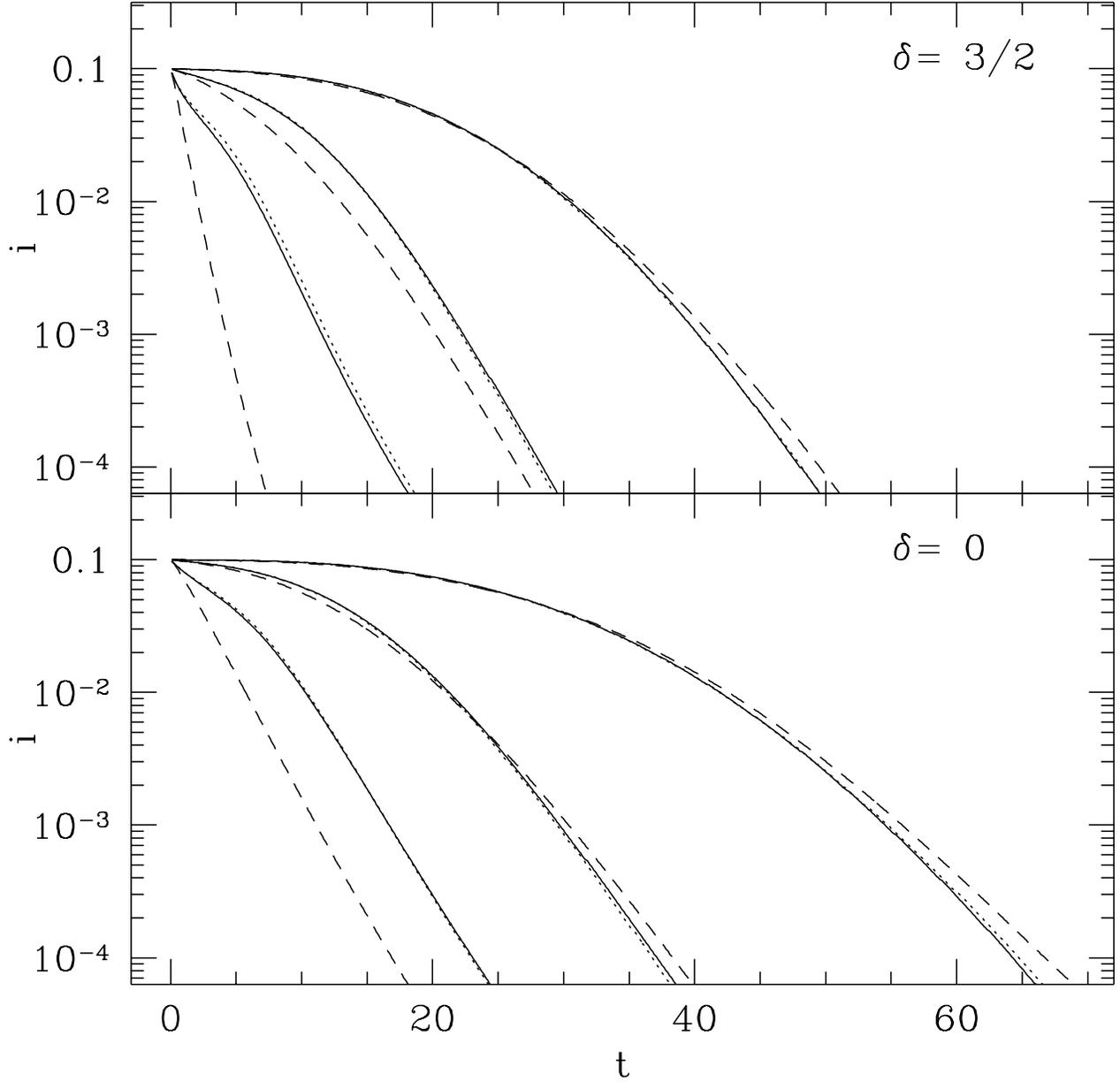}
\caption{Damping of the disk inclination $i$ at $R=2 R_{\rm in}$ for varying disk 
         viscosity. Upper panel is for $\delta = 3/2$, lower panel is
         for $\delta = 0$. Solid curves are for $\nu_1 / \nu_2 = 1$, dashed 
         lines are for $\nu_1 / \nu_2 = 5$, dotted lines are for 
         $\nu_1 / \nu_2 = 0.2$. In each panel, the upper family of curves are
         for $\nu_{20} = 10^{-4}$, middle for $\nu_{20} = 10^{-3}$, and the 
         fastest decaying curves are for $\nu_{20} = 10^{-2}$. The time units 
         are such that $\vert {\bf \Omega_p} \vert^{-1} = 1$ at $R=2$.}
         
\end{figure}  

\begin{figure}
\plotone{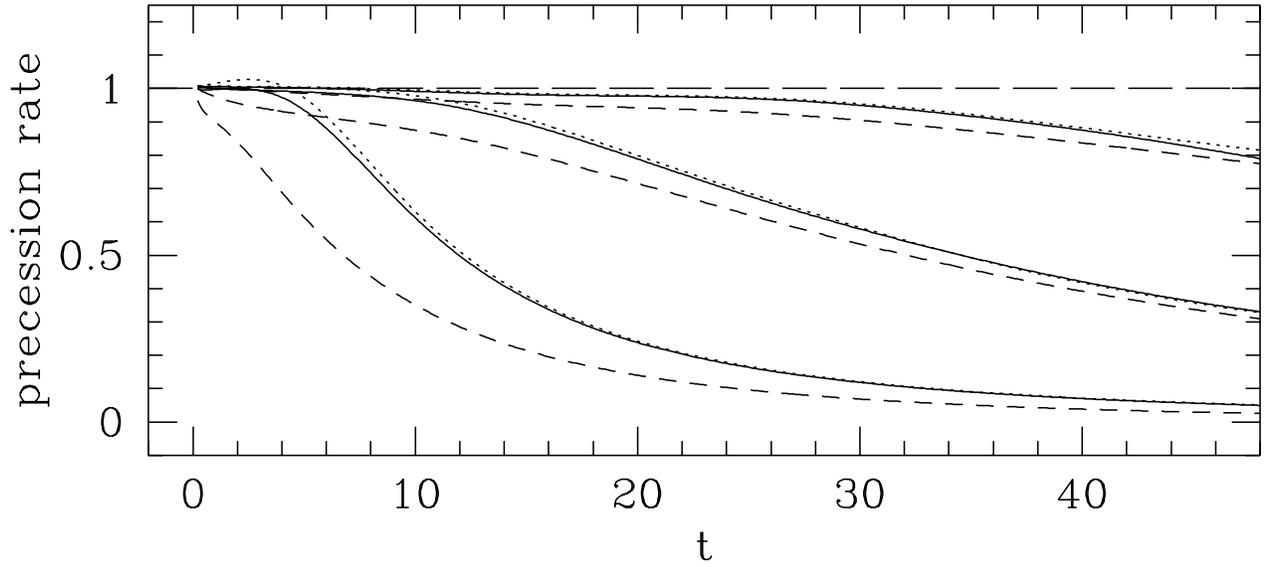}
\caption{Precession rate at $R=2$ for the models with $\delta = 3/2$. Lines
         correspond to the same models as for Fig.~3. The long dashed line 
         shows the predicted Lense-Thirring precession rate at this radius.
         Note that the time axis is reduced as compared to Fig.~3.}
\end{figure}  	

\begin{figure}
\plotone{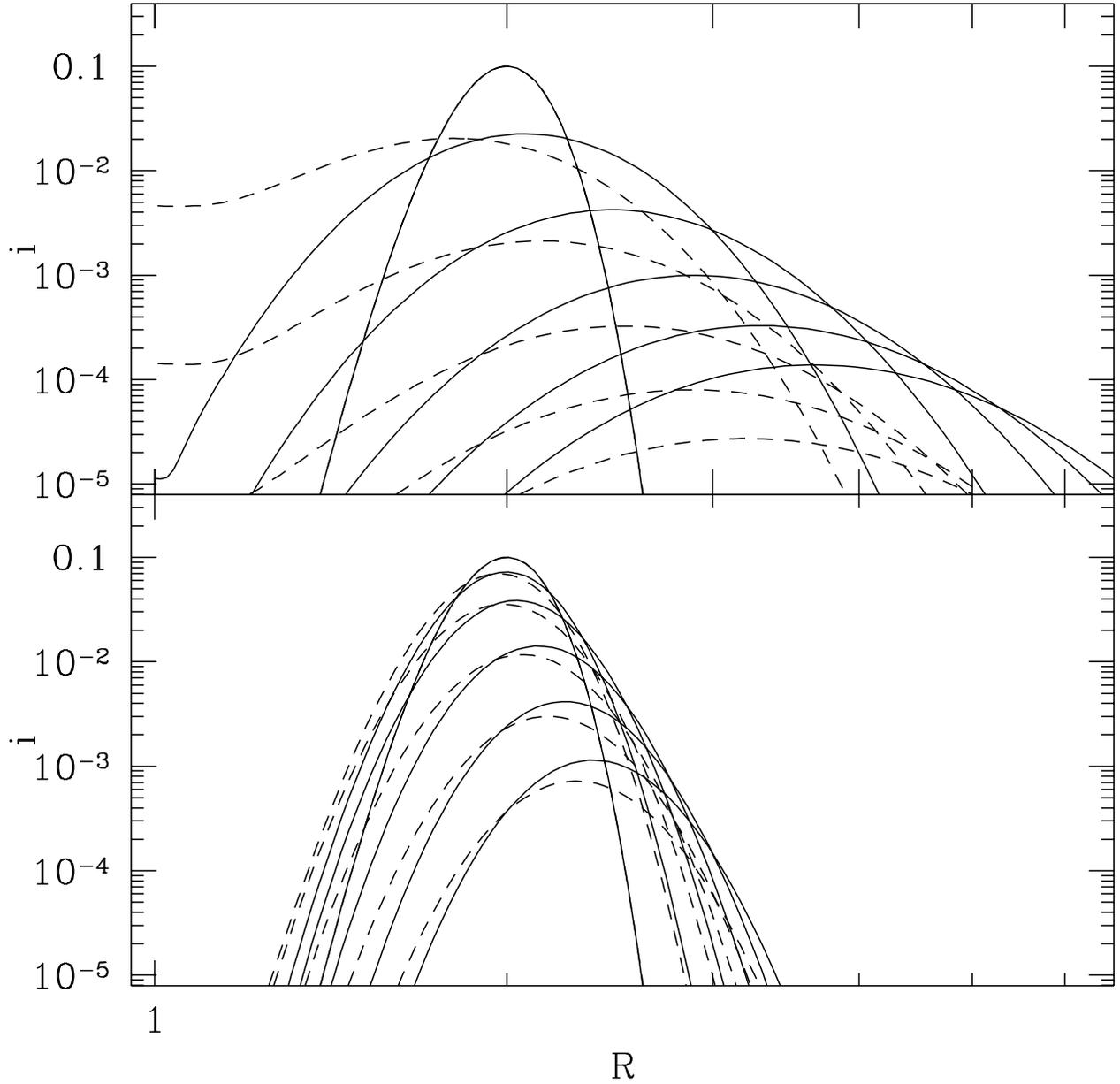}
\caption{Decay of the warp calculated using the linearized twist equations 
         (solid lines) compared to the numerical solution of the full 
         equations (dashed lines). The upper panel is for $\delta = 3/2$, 
         $\nu_{10} = \nu_{20} = 10^{-2}$, the lower panel is for the 
         same parameters except that
         $\nu_{10} = \nu_{20} = 10^{-3}$.}
\end{figure}  

\begin{figure}
\plotone{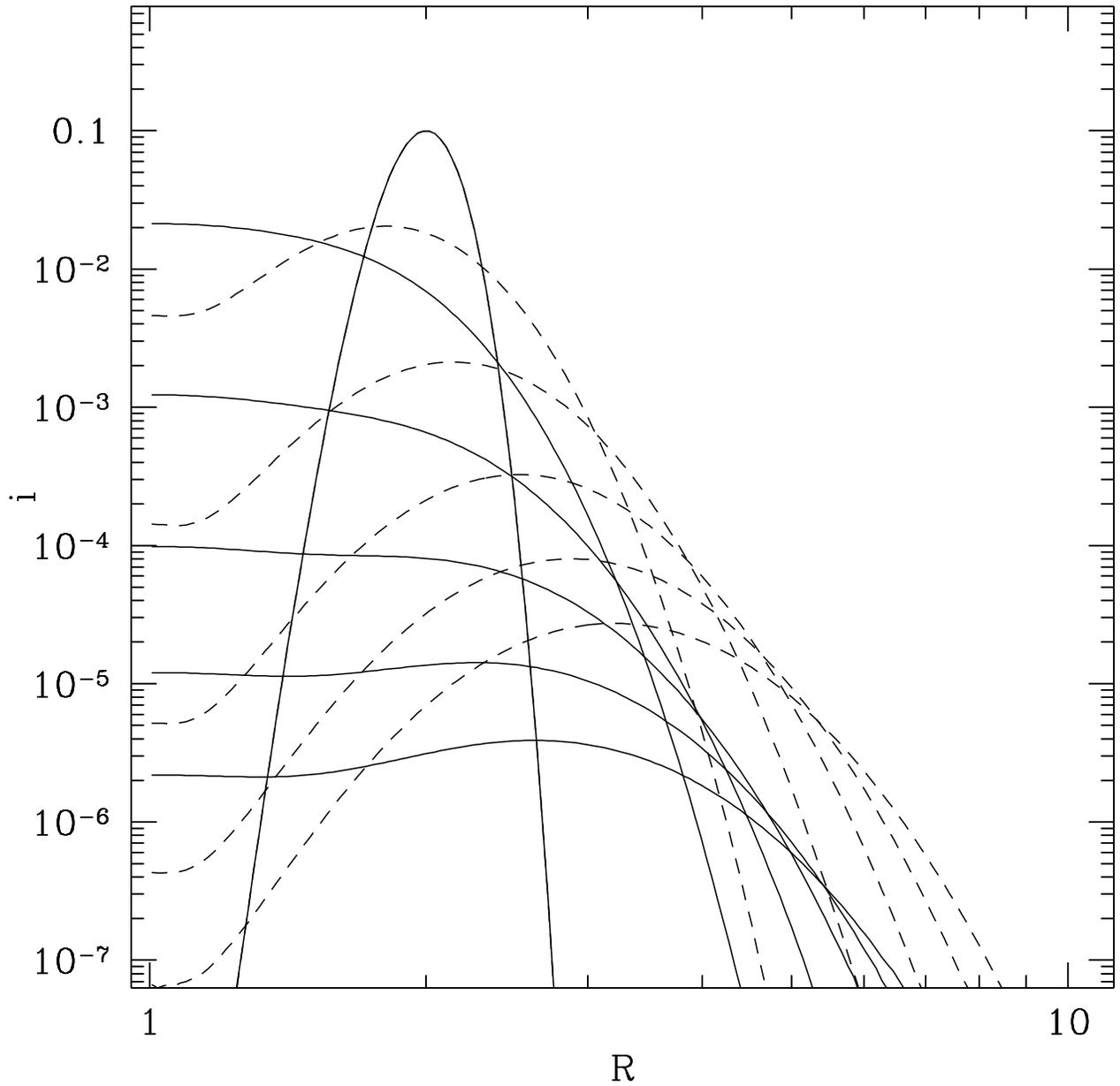}
\caption{Comparison of calculations using a Newtonian point mass potential 
         (dashed lines) with a Paczynski-Wiita potential (solid lines). 
         The parameters are as for Fig.~1. There are differences in the 
         shape of the warp at late times, but the damping rates are 
         almost identical.}
\end{figure} 

\end{document}